\documentclass[a4paper,11pt]{article}
\usepackage{jheppub}

\usepackage{makecell}
\usepackage{float}
\usepackage{color}
\usepackage{ulem}

\title{\boldmath Towards early dark energy and $n_s$=1 with Planck, ACT and SPT}

\author[a]{Jun-Qian Jiang}
\author[a,b,c,d]{Yun-Song Piao}

\affiliation[a]{School of Physics, University of Chinese Academy
of Sciences, Beijing 100049, China}

\affiliation[b]{School of Fundamental Physics and Mathematical
Sciences, Hangzhou Institute for Advanced Study, UCAS, Hangzhou
310024, China}

\affiliation[c]{International Center for Theoretical Physics
Asia-Pacific, Beijing/Hangzhou, China}

\affiliation[d]{Institute of Theoretical Physics, Chinese Academy
of Sciences, P.O. Box 2735, Beijing 100190, China}

\emailAdd{jqjiang@zju.edu.cn} \emailAdd{yspiao@ucas.ac.cn}

\abstract{ We investigate the constraints on early dark energy
(EDE) by combining the most recent CMB observations available, ACT
DR4, SPT-3G, and Planck2018 ($\ell_\text{TT,max}=1000$) data. 
This
combined CMB dataset favors non-zero EDE fractions and large Hubble
constants. The inclusion of BAO+Pantheon data has little effect on the 
results, leads to $H_0=71.6( 72.9)_{-1.5}^{+2.0}$ and
$73.17(72.74)^{+0.55}_{-0.77}$ km/s/Mpc for axion-like EDE and
AdS-EDE, respectively. The axion-like EDE can fit the data significantly better ($\Delta \chi^2 \lesssim -10$) than
$\Lambda$CDM, which is mainly driven by the ACT data. It is found again that if the current $H_0$ 
measured locally is correct, complete resolution of the Hubble tension seems to be pointing towards a 
scale invariant Harrison-Zeldovich spectrum of primordial scalar perturbation, 
i.e. $n_s=1$ for $H_0\sim 73$ km/s/Mpc.}

\begin{document}
\maketitle
\flushbottom

\section{Introduction}
\label{sec:1}

As the precision of cosmological observations increases, the
standard $\Lambda$CDM model is facing challenges. One of the
critical problems is the $4 \sim 6 \sigma$ tension between the
Hubble constant based on the early universe with the $\Lambda$CDM
model and that measured by direct observations of the local
universe without assuming the $\Lambda$CDM model
\cite{Verde:2019ivm,Riess:2019qba} (see e.g.
\cite{DiValentino:2020zio,DiValentino:2021izs,Perivolaropoulos:2021jda}
for recent reviews), dubbed as the Hubble tension.
The systematic errors are unable to explain it entirely, so
modifications to the cosmological model are required
\cite{DiValentino:2016hlg,Mortsell:2018mfj,Vagnozzi:2019ezj,Knox:2019rjx,Schoneberg:2021qvd}.

Pre-recombination early dark energy (EDE)
\cite{Karwal:2016vyq,Poulin:2018cxd} is one of the most promising
route to resolve the Hubble tension. In corresponding scenario,
energy injection before recombination led to faster expansion of
the Universe, which so reduced the sound horizon. As
$\theta^*_\text{s}=r^*_\text{s}/D^*_A$ is measured precisely by
CMB observations, we can obtain a higher $H_0$ while keeping
late-time physics unchanged. There are various kinds of
phenomenological models
\cite{Poulin:2018cxd,Kaloper:2019lpl,Agrawal:2019lmo,Lin:2019qug,Smith:2019ihp,Niedermann:2019olb,Sakstein:2019fmf,Ye:2020btb,Gogoi:2020qif,Braglia:2020bym,Lin:2020jcb,Seto:2021xua,Nojiri:2021dze,Karwal:2021vpk}
for this early energy injection with effective fluids or scalar
fields, see also \cite{Zumalacarregui:2020cjh,Ballesteros:2020sik,Braglia:2020auw}. Decay of EDE must be rapid so as not to spoil other
observations, which can be achieved by an oscillatory potential in
axion-like EDE, e.g.Refs.\cite{Poulin:2018cxd,Smith:2019ihp}, or
by an anti-de Sitter (AdS) phase in AdS-EDE \cite{Ye:2020btb}.
Both showed a better fit to the CMB,
baryon acoustic oscillation (BAO) and local $H_0$ data.

Planck data, the most precise large-scale CMB observation
currently available, alone seems not favor axion-like EDE model
(see Ref.\cite{Hill:2020osr}). However, the Planck data itself is
debatable, especially its small scale part of TT power spectrum.
The inconsistency between the $\ell<1000$ and $\ell>1000$ part of
Planck's TT power spectrum has been pointed out in
Refs.\cite{Addison:2015wyg,Planck:2016tof}. Moreover, the
smoothing effect of gravitational lensing on acoustic peaks of the
CMB power spectrum exceeds that expected in $\Lambda$CDM model
\cite{Addison:2015wyg,Motloch:2019gux}. However, ground-based CMB
observations, such as ACT and SPT, providing precise
measurements on small scale power spectrum have not found
this over-smoothing effect
\cite{SPT:2017jdf,ACT:2020gnv,SPT-3G:2021eoc}.

It has been found that, without the small scale part of Planck TT
power spectrum, a large fraction of EDE and a large Hubble
constant are possible. Recently, combined analysis of Planck data
($\ell_\mathrm{TT} \lesssim 1000$) with ACT or SPT data have been performed for
EDE models, such as Planck + SPTpol for power-law potential EDE
\cite{Chudaykin:2020acu}, axion-like EDE \cite{Chudaykin:2020igl},
AdS-EDE \cite{Jiang:2021bab} and Planck + ACT DR4 for axion-like
EDE \cite{Hill:2021yec,Poulin:2021bjr} and NEDE
\cite{Poulin:2021bjr}. And also Planck + ACT DR4 + SPT-3G Y1 for
axion-like EDE \cite{LaPosta:2021pgm}.

In this work, in view of the important role of ground-based CMB
observations, we investigate the constraints on axion-like EDE and
AdS-EDE models using the combination of Planck18 data (we exclude
$\ell>1000$ part of Planck TT power spectrum) with the recent
SPT-3G Y1 and ACT DR4 data, with and without BAO and Pantheon
data. We find that this combined CMB dataset favors a non-zero EDE
fraction and a large Hubble constant for both models. In
Ref.\cite{Ye:2021nej}, it has been found that with
fullPlanck+BAO+Pantheon dataset the pre-recombination solutions of
the Hubble tension implies a scale-invariant Harrison-Zeldovich
spectrum of primordial scalar perturbation, i.e. $n_s= 1$ for
$H_0\sim 73$ km/s/Mpc. It is also interesting to recheck this
conclusion with our combined CMB dataset.

The paper is outlined as follows. We explain our data, models and
methodology in \autoref{sec:2}. Results are presented in
\autoref{sec:3}. Then we analyse and discuss it in
\autoref{sec:4}. Finally, we conclude in \autoref{sec:5}.

\section{Model, Data and Methodology}
\label{sec:2}

The first EDE model is: axion-like EDE, where the energy injection
before recombination is achieved by a scale field with axion-like
potential \cite{Poulin:2018cxd}:
\begin{equation}
    V( \phi ) = m^2 f^2 ( 1-\cos\theta)^{n} \text{, where } \theta=\phi /f \in [-\pi, \pi]
\end{equation}
It describes the axion for $n=1$, which naturally arises in high
energy theory. Initially, the field sit in the upper region of its
potential due to the Hubble fiction, resulted in $w \approx -1$,
i.e. an dark energy injection. Afterwards it will roll to the
bottom of the potential and oscillate with $w \approx
(n-1)/(n+1)$. Therefore, the energy will redshift faster than
matter if $n>1$, avoid degrading other measurements (e.g. matter
density and CMB).
Here, we fixed $n=3$, which is a suitable value for the current data \cite{Smith:2019ihp}.

Another model we consider is AdS-EDE, with phenomenological
potential \cite{Ye:2020btb,Jiang:2021bab}: \footnote{Other
potentials are also possible \cite{Ye:2020oix}.}
\begin{equation}
    V(\phi)=\left\{\begin{array}{ll}
    V_{0}\left(\dfrac{\phi}{M_{\text{Pl}}}\right)^{4}-V_{\text{AdS}} & ,\quad \dfrac{\phi}{M_{\text{Pl}}}<\left(\dfrac{V_{\text{AdS}}}{V_{0}}\right)^{1 / 4} \\
    0 &, \quad\dfrac{\phi}{M_{\text{Pl}}}>\left(\dfrac{V_\text{AdS}}{V_{0}}\right)^{1 / 4}
    \end{array}\right.
\end{equation}
where $M_{\text{Pl}}$ is the reduced Planck mass. $V_{\text{AdS}}$
is the depth of AdS phase. The significant difference from
axion-like EDE is that the energy redshifts in an AdS phase. In an
AdS phase with $w> 1$ the energy of EDE can redshift faster than
that in oscillation phase, thus results less destruction to other
measurement. \footnote{The difference between the power-law and
cosine potentials in the axion EDE is also important
\cite{Smith:2019ihp}.} It is well-known that AdS vacua is
ubiquitous in high energy theories, so the AdS-EDE model can be
well-motivated, see also the applications of AdS vacua
to late Universe
\cite{Akarsu:2019hmw,Visinelli:2019qqu,Dutta:2018vmq,Calderon:2020hoc,Ruchika:2020avj,Akarsu:2021fol,Sen:2021wld}.

We consider the following CMB data sets at first:
\begin{itemize}
    \item \textbf{Planck 2018}: We use the low-$\ell$ TT,EE \texttt{Commander} likeihoods and high-$\ell$ TT,TE,EE \texttt{Plik} likeihoods, with also the reconstructed lensing power spectrum \cite{Planck:2018vyg}.
    \item \textbf{SPT-3G Y1}: We use the public SPT-3G likeihood\footnote{\url{https://github.com/SouthPoleTelescope/spt3g\_y1\_dist}}, which includes TE and EE power spectrum within multipoles $300<\ell<3000$ \cite{SPT-3G:2021eoc}.
    \item \textbf{ACT DR4}: We use the marginalized likeihood\footnote{\url{https://github.com/ACTCollaboration/pyactlike}} from ACT
    Data Release 4, which includes TE and EE power spectrum within multipoles $326<\ell<4325$ and TT power spectrum within multipoles $576<\ell<4325$ \cite{ACT:2020frw}.
\end{itemize}
This data set combination is confirmed in
Ref.\cite{SPT-3G:2021wgf}, which showed $H_0=67.49\pm0.53$
km/s/Mpc for $\Lambda$CDM model. As mentioned, it might be better
to discard the small scale part of Planck's TT, so when combine
Planck measurement with ACT and SPT data, we cut the Planck's
high-$\ell$ TT power spectrum to $\ell_\text{TT,max} = 1000$.
\footnote{This choice of $\ell_\text{TT,max}$ is the same as
Refs.\cite{Chudaykin:2020acu,Chudaykin:2020igl,Jiang:2021bab} and
close to Refs.\cite{Poulin:2021bjr} (where $\ell_\text{TT,max} =
1060$)
 and \cite{Hill:2021yec,LaPosta:2021pgm} (where $\ell_\text{TT,max} = 650$). }
Then, we use galaxy BAO measurements from 6DF
\cite{Beutler:2011hx}, SDSS DR7 MGS \cite{Ross:2014qpa} in low-$z$
and BOSS DR12 \cite{BOSS:2016wmc} in high-$z$. Type Ia supernovae
from Pantheon \cite{Scolnic:2017caz} is also used.

We perform MCMC sampling with \texttt{Cobaya}
\cite{Torrado:2020dgo}. The models are calculated using the
modified \texttt{CLASS} \cite{Blas:2011rf}
\footnote{The codes are available at \url{https://github.com/PoulinV/AxiCLASS} for axion-like EDE and \url{https://github.com/genye00/class\_multiscf} for AdS-EDE.}
, where we improved the
accuracy for the calculation of the lensing effect since it has
non-negligible effects on the small-scale CMB power spectrum. The
Gelman-Rubin criterion for all chains is converged to $R-1<0.1$.

Here, we adopt same parameters and priors as
Refs.\cite{Poulin:2021bjr} and \cite{Jiang:2021bab},
for axion-like EDE model: $\log_{10}(z_c)\in[2, 4.5]$,
$f_\text{EDE}\in[0, 0.3]$, where $z_c$ is the redshift
at which the field starts rolling and $f_\text{EDE}$ is the energy
fraction of EDE at $z_c$, the initial position of EDE field
$\Theta_\text{ini}\in[0, 3.1]$, while for AdS-EDE model:
$\ln(1+z_c)\in[7.5, 9]$, $f_\text{EDE}\in[0, 0.3]$. In order to
have a significant AdS phase while make the field
able to climb out of the AdS well, we fixed
$\alpha_{\text{AdS}} \equiv
\left(\rho_\text{m}(z_c)+\rho_\text{r}(z_c)\right) / V_\text{AdS}
=3.79 \times 10^{-4}$ as \cite{Ye:2020btb}. The neutrino
assumption is the same as Planck \cite{Planck:2018vyg}. The
posterior distribution is plotted using \texttt{GetDist}
\cite{Lewis:2019xzd}.
The bestfit points are obtained through \texttt{BOBYQA}
\cite{cartis2019improving,cartis2018escaping,powell2009bobyqa}.

\section{Results}
\label{sec:3}

\subsection{axion-like EDE}

\begin{figure}[tbp]
\includegraphics[width=.9\textwidth]{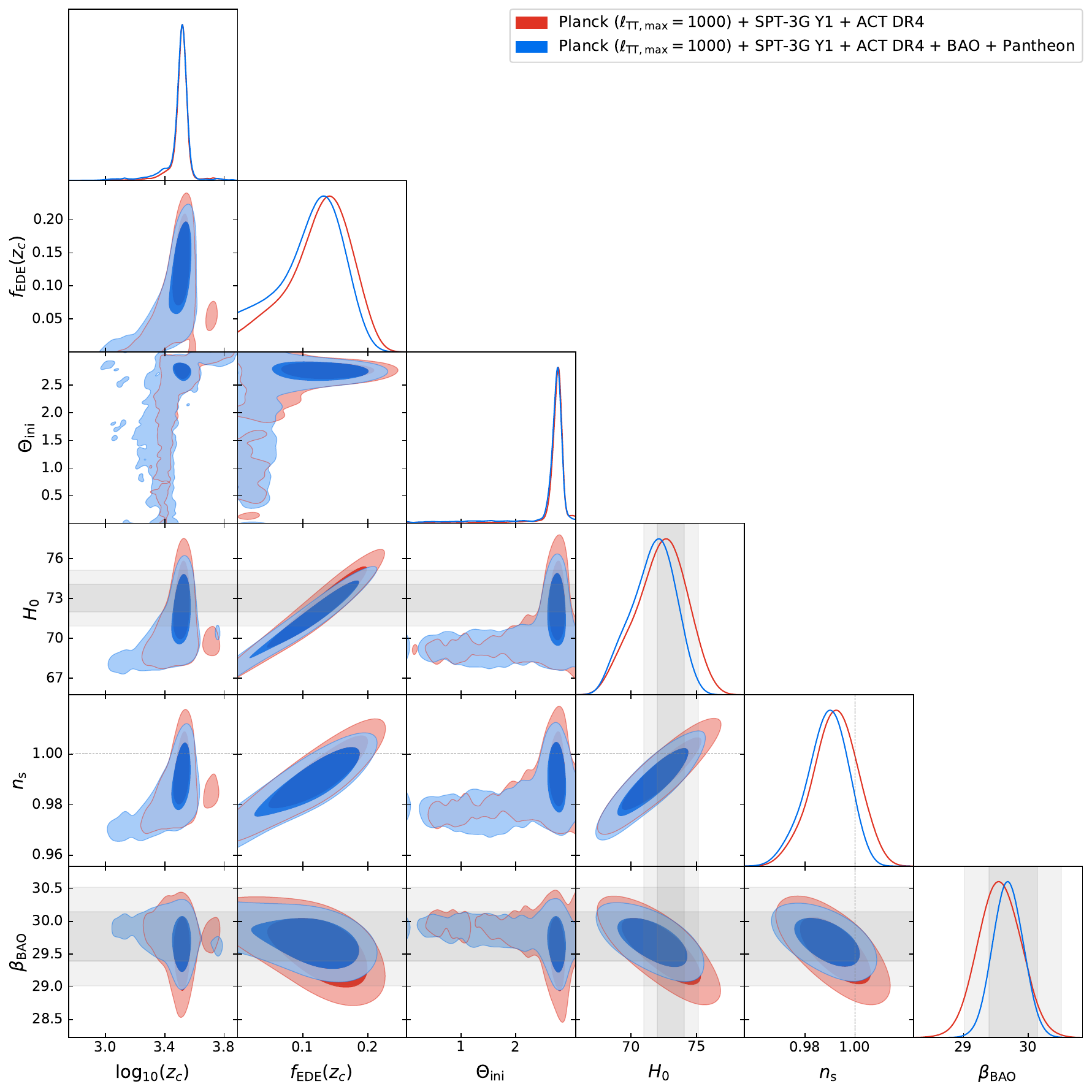}
\caption{\label{fig:axionEDE} Posterior distributions of relevant
parameters in axion-like EDE model (68\% and 95\% confidence
range). Grey bands represent the 1$\sigma$ and 2$\sigma$ regions
of the SH0ES measurement \cite{Riess:2021jrx} and
model-independent constraint on $\beta_\text{BAO}$ from BAO and
SN, respectively.}
\end{figure}

\begin{table}[]
    \centering
\begin{tabular}{|c|c|c|c|}
\hline
 parameters & \makecell[c]{Planck($\ell _{\text{TT, max}} =1000$)\\+ACT DR4+SPT-3G Y1} & \makecell[c]{Planck($\ell _{\text{TT, max}} =1000$)\\+ACT DR4+SPT-3G Y1\\+BAO+Pantheon} & \makecell[c]{Planck\\+BAO+Pantheon} \\
\hline
\hline
 $f_{\text{EDE}}$ & $0.127( 0.150)^{+0.058}_{-0.034}$ & $0.112( 0.148)_{-0.038}^{+0.063}$ & $< 0.084( 0.09)$ \\
 $\log_{10}( z_{c})$ & $3.507( 3.514)^{+0.046}_{-0.024}$ & $3.495( 3.521)_{-0.023}^{+0.064}$ & unconstrained $( 3.569)$ \\
 $\Theta _{\text{ini}}$ & $2.66( 2.80)^{+0.21}_{+0.020}$ & $2.53( 2.75)_{+0.11}^{+0.35}$ & $1.933( 2.773)_{-0.44}^{+1.2}$ \\
 $H_{0}$ & $72.4(73.4)^{+2.2}_{-1.7}$ & $71.6( 72.9)_{-1.5}^{+2.0}$ & $68.6( 70.88)_{-1.1}^{+0.55}$ \\
 $100\omega _{\text{b}}$ & $2.263( 2.262)_{-0.019}^{+0.017}$ & $2.260( 2.267)_{-0.020}^{+0.017}$ & $2.257( 2.270)_{-0.02}^{+0.017}$ \\
 $\omega _{\text{cdm}}$ & $0.1320( 0.1340)_{-0.0050}^{+0.0062}$ & $0.1307( 0.1348)_{-0.0053}^{+0.0067}$ & $0.1219( 0.1278)_{-0.0034}^{+0.0013}$ \\
 $10^{9} A_{\text{s}}$ & $2.139( 2.141) \pm 0.035$ & $2.126( 2.140) \pm 0.035$ & $2.118( 2.159)_{-0.034}^{+0.031}$ \\
 $n_{\text{s}}$ & $0.9923( 0.9962)_{-0.0086}^{+0.0098}$ & $0.9885( 0.9933)_{-0.0078}^{+0.0094}$ & $0.9719( 0.9850)_{-0.0076}^{+0.0048}$ \\
 $\tau _{\text{reio}}$ & $0.0531( 0.0526) \pm 0.0073$ & $0.0509( 0.0516)_{-0.0070}^{+0.0078}$ & $0.0569( 0.0617)_{-0.0078}^{+0.0071}$ \\
\hline
 $S_{8}$ & $0.832( 0.830) \pm 0.014$ & $0.833( 0.839) \pm 0.014$ & $0.828( 0.836) \pm 0.013$ \\
 $\Omega _{\text{m}}$ & $0.2964( 0.2921) \pm 0.0085$ & $0.3000( 0.2975) \pm 0.0058$ & $0.3085( 0.3008) \pm 0.0059$ \\
 \hline
    \end{tabular}
\caption{The mean (best-fit) $\pm 1 \sigma$ errors of parameters
in axion-like EDE model for each dataset combination, The result
with fullPlanck+BAO+Pantheon is from Ref.\cite{Poulin:2021bjr}.}
    \label{tab:axionEDE}
\end{table}

We show the posterior distribution in \autoref{fig:axionEDE} and
the mean (best-fit) values of cosmological parameters in
\autoref{tab:axionEDE}. The result is similar to
Refs.\cite{2112.10754} with $\ell_\text{TT,max}=650$, where BAO
and SN data were not included. We see that Planck($\ell
_{\text{TT, max}} =1000$)+ACT DR4+SPT-3G Y1 dataset alone favor a
non-zero fraction of EDE $f_\text{EDE} = 0.127( 0.150)^{+0.058}_{-0.034}$
and a large Hubble constant $H_0 =
72.4(73.4)^{+2.2}_{-1.7}$, which is compatible with the SH0ES
result \cite{Riess:2021jrx}.

The inclusion of BAO+Pantheon does not alter the result too much.
This can be confirmed by checking $\beta_{\text{ BAO}} \equiv
c/\left(H r_s^{\text{darg}}\right)$, which BAO+Pantheon mainly
constrain. We can find in \autoref{fig:axionEDE} that the
$\beta_{\text{ BAO}}$ obtained from CMB data alone is consistent
with that constrained by BAO+SN observations.
\footnote{BAO+Pantheon constraints on $\beta_{\text{ BAO}}$ are
calculated in \cite{Jiang:2021bab} based on the model-independent
approach described in \cite{Bernal:2016gxb,Aylor:2018drw}.}
Besides, the preference for large $\Theta_\text{ini}$ due to the
inclusion of Placnk's high-$\ell$ TE,EE power spectrum, is in
agreement with the analysis in \cite{Smith:2019ihp}. Unlike
Ref.\cite{Hill:2021yec}, we find $\log_{10}(z_c) \approx 3.5$,
close to the matter-radiation equality, even with lensing and BAO
data included. This is because a larger $\Theta_\text{ini}$
prefers the parameter region with smaller $z_c$, which can also be
found in the Appendix B of Ref.\cite{Poulin:2021bjr}.

We present the $\chi^2$ for the bestfit points in
\autoref{tab:chi2_fPlSA} and \autoref{tab:chi2_fPlSABP}. We found
significant improvements $\Delta \chi^2 \approx -11$ without and
with BAO+Pantheon for axion-like EDE model compared to
$\Lambda$CDM in fitting CMB data. The main improvement is come
from CMB data, especially ACT DR4 and Planck high-$\ell $ part.

\subsection{AdS-EDE}

\begin{figure}[tbp]
\includegraphics[width=.9\textwidth]{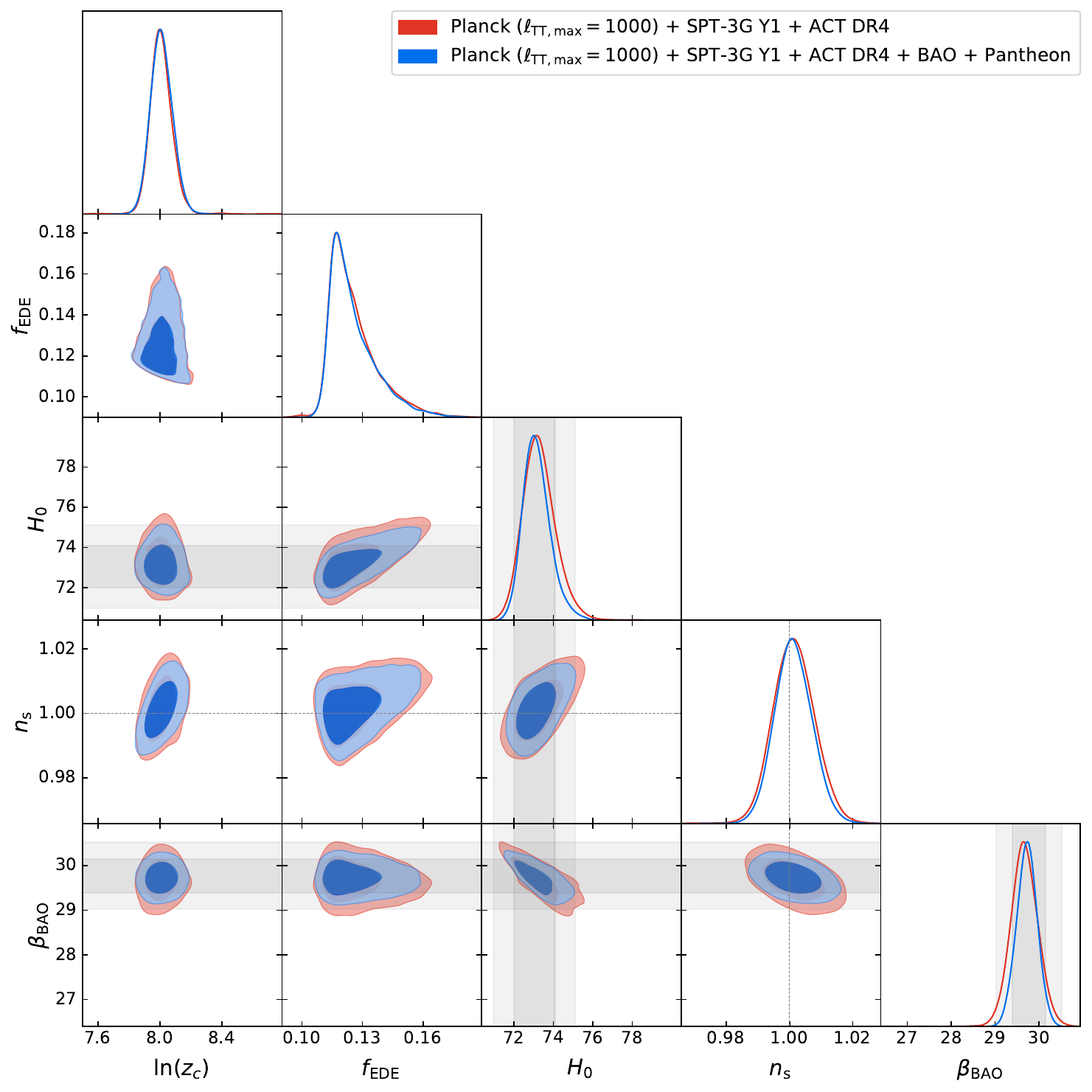}
\caption{\label{fig:AdSEDE} Posterior distributions of relevant
parameters in AdS-EDE model (68\% and 95\% confidence range). Grey
bands represent the 1$\sigma$ and 2$\sigma$ regions of the SH0ES
measurement \cite{Riess:2021jrx} and model-independent constraint
on $\beta_\text{BAO}$ from BAO and SN, respectively.}
\end{figure}

\begin{table}[]
    \centering
\begin{tabular}{|c|c|c|c|}
\hline
 parameters & \makecell[c]{Planck($\ell _{\text{TT, max}} =1000$)\\+ACT DR4+SPT-3G Y1} & \makecell[c]{Planck($\ell _{\text{TT, max}} =1000$)\\+ACT DR4+SPT-3G Y1\\+BAO+Pantheon} & \makecell[c]{Planck\\+BAO+Pantheon} \\
\hline
\hline
 $f _{\text{EDE}}$ & $0.1257(0.1199)^{+0.0050}_{-0.014}$ & $0.1253( 0.1163)^{+0.0049}_{-0.013}$ & $0.1124( 0.1084)_{-0.0070}^{+0.0038}$ \\
 $\ln( 1+z_{c})$ & $8.009(8.007)^{+0.060}_{-0.074}$ & $8.011( 7.968)^{+0.065}_{-0.074}$ & $8.153( 8.147)_{-0.084}^{+0.075}$ \\
 $H_{0}$ & $73.31(72.98)^{+0.68}_{-0.93}$ & $73.17(72.74)^{+0.55}_{-0.77}$ & $72.52( 72.46) \pm 0.51$ \\
 $100\omega _{\text{b}}$ & $2.309(2.311)\pm 0.021$ & $2.308(2.299)\pm 0.020$ & $2.341( 2.331)_{-0.016}^{+0.018}$ \\
 $\omega _{\text{cdm}}$ & $0.1367(0.1361)^{+0.0021}_{-0.0028}$ & $0.1370( 0.1358)^{+0.0016}_{-0.0026}$ & $0.1346( 0.1336)_{-0.0018}^{+0.0016}$ \\
 $10^{9} A_{\text{s}}$ & $2.134(2.138)^{+0.040}_{-0.033}$ & $2.13e( 2.120)^{+0.035}_{-0.031}$ & $2.175( 2.159) \pm 0.033$ \\
 $n_{\text{s}}$ & $1.0014(1.0021)\pm 0.0067$ & $1.0013( 0.9989)\pm 0.0059$ & $0.9964( 0.9949)_{-0.0041}^{+0.0047}$ \\
 $\tau _{\text{reio}}$ & $0.0453(0.0468)^{+0.010}_{-0.0082}$ & $0.0449( 0.0433)^{+0.0088}_{-0.0075}$ & $0.0545( 0.0523)_{-0.0079}^{+0.0071}$ \\
\hline
 $S_{8}$ & $0.857(0.860)\pm 0.017$ & $0.860(0.858)\pm 0.012$ & $0.863( 0.856) \pm 0.011$ \\
 $\Omega _{\text{m}}$ & $0.2987(0.3001)\pm 0.0087$ & $0.3002(0.3014)\pm 0.0060$ & $0.3016( 0.3002) \pm 0.0051$ \\
 \hline
\end{tabular}
\caption{The mean (best-fit) $\pm 1 \sigma$ errors of parameters
in AdS-EDE model for each dataset combination, The result with
fullPlanck+BAO+Pantheon is from Ref.\cite{Jiang:2021bab}.}
    \label{tab:AdSEDE}
\end{table}

The combined CMB dataset still favor a non-zero fraction of
AdS-EDE $f_\text{EDE} = 0.1257(0.1199)^{+0.0050}_{-0.014}$ and a
large Hubble constant $H_0 = 73.31(72.98)^{+0.68}_{-0.93}$, see
the posterior distribution in \autoref{fig:AdSEDE} and the mean
(best-fit) values in \autoref{tab:AdSEDE}, and the inclusion of
BAO+Pantheon does not change the results. However, different from
that for axion-like EDE, the fullPlanck+BAO+Pantheon dataset still favor
a large $H_0 = 72.52(72.46) \pm 0.51$ for AdS-EDE.
A distinct non-Gaussian distribution
is shown in $f_\text{EDE}$-$\ln(z_c)$ plane. This is a reflection of
AdS bound, or else the EDE field will be unable to climb out of AdS well.
These results are consistent with out previous results \cite{Jiang:2021bab},
where Planck ($\ell_{\text{TT, max}} =1000$) and SPTpol were included as CMB datasets.
Furthermore, we now have smaller error bars, as SPT-3G Y1 data has stronger constraining power than SPTpol, and ACT DR4 has comparable constraining power.

We present the $\chi^2$ for the bestfit points in
\autoref{tab:chi2_fPlSA} and \autoref{tab:chi2_fPlSABP}. We find
that AdS-EDE shows about $\Delta \chi^2 \approx +2$ worse fit to
the corresponding dataset than $\Lambda$CDM, mainly comes from
Planck lensing. This difference is statistically insignificant.
However, similar to $\Lambda$CDM, AdS-EDE does not fit better to
Planck high-$\ell$ part and ACT data, compared with axion-like
EDE. Here, we only have investigated the simplest AdS-EDE model,
actually other AdS potentials are also possible. Note that the
bestfit point is so close to AdS bound that the ``real'' bestfit
point might be covered by it. The AdS bound is controlled by
$\alpha_{\text{AdS}}$, thus a smaller $\alpha_{\text{AdS}}$ can
bring a better fit.

\begin{table}[!h]
        \centering
\begin{tabular}{|c|c|c|c|}
\hline
  & $\Lambda $CDM & axion-like EDE & AdS-EDE \\
\hline
 Planck low-$\ell $ TT & 21.74 & 20.63 & 20.32 \\
 Planck low-$\ell $ EE & 395.72 & 395.83 & 396.21 \\
 Planck high-$\ell $ TTTEEE ($\ell _{\text{TT, max}} =1000$) & 1985.91 & 1983.32 & 1982.90 \\
 Planck lensing & 9.16 & 10.73 & 11.32 \\
 ACT DR4 & 297.48 & 289.17 & 300.03 \\
 SPT-3G Y1 & 1117.31 & 1116.94 & 1118.93 \\
\hline
 total $\chi ^{2}$ & 3827.32 & 3816.61 & 3829.72 \\
\hline
 $\chi _{\text{model} -\Lambda \text{CDM}}^{2}$ &  & $-10.71$ & +2.40 \\
 \hline
\end{tabular}
\caption{$\chi^2$ values for best-fit models to Planck ($\ell
_{\text{TT, max}} =1000$) + ACT DR4 + SPT-3G Y1 dataset.}
    \label{tab:chi2_fPlSA}
\end{table}

\begin{table}[!h]
        \centering
\begin{tabular}{|c|c|c|c|}
\hline
  & $\Lambda $CDM & axion-like EDE & AdS-EDE \\
\hline
 Planck low-$\ell $ TT & 21.54 & 20.81 & 20.45 \\
 Planck low-$\ell $ EE & 396.11 & 395.84 & 395.85 \\
 Planck high-$\ell $ TTTEEE ($\ell _{\text{TT, max}} =1000$) & 1986.18 & 1981.81 & 1984.92 \\
 Planck lensing & 9.27 & 10.41 & 11.20 \\
 ACT DR4 & 297.60 & 290.54 & 297.92 \\
 SPT-3G Y1 & 1117.60 & 1117.01 & 1118.44 \\
 BAO low-$z$ & 1.67 & 2.25 & 1.86 \\
 BOSS DR12 BAO & 3.57 & 3.50 & 3.50 \\
 Pantheon & 1034.81 & 1034.74 & 1034.75 \\
\hline
 total $\chi ^{2}$ & 4868.34 & 4856.90 & 4869.91 \\
\hline
 $\chi _{\text{model} -\Lambda \text{CDM}}^{2}$ &  & $-11.44$ & +1.57 \\
 \hline
\end{tabular}
\caption{$\chi^2$ values for best-fit models to Planck ($\ell
_{\text{TT, max}} =1000$) + ACT DR4 + SPT-3G Y1 + BAO + Pantheon
dataset.}
    \label{tab:chi2_fPlSABP}
\end{table}

\section{Discussion}
\label{sec:4}

\begin{figure}[tbp]
\centering
\includegraphics[width=\textwidth]{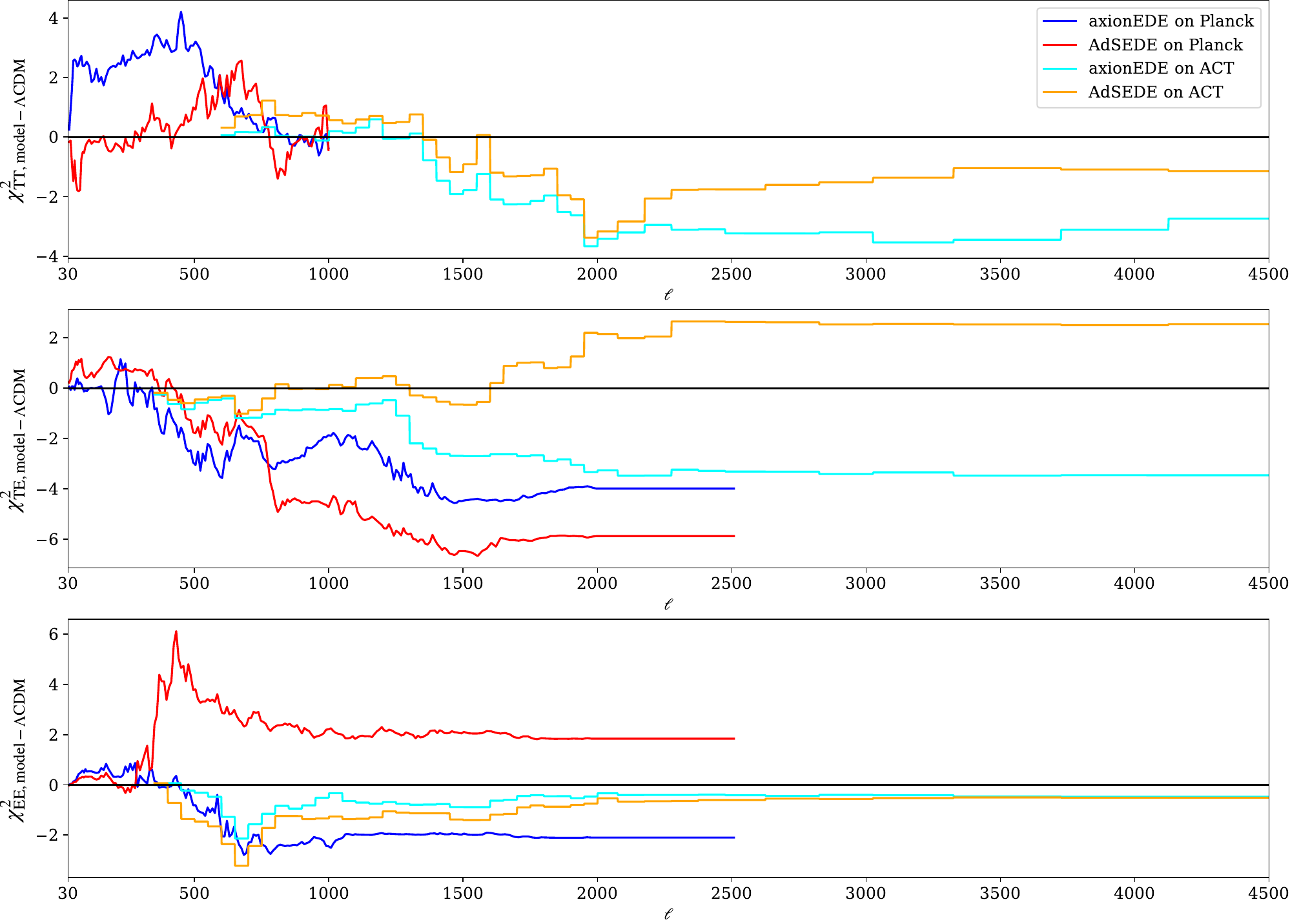}
\caption{\label{fig:chi2} The cumulative
$\chi^2_{\mathrm{model}-\Lambda\mathrm{CDM}}$ of each part of data
in the best-fit points to Planck ($\ell _{\text{TT, max}} =1000$)
+ ACT DR4 + SPT-3G Y1 + BAO + Pantheon.}
\end{figure}

\begin{figure}[tbp]
\centering
\includegraphics[width=\textwidth]{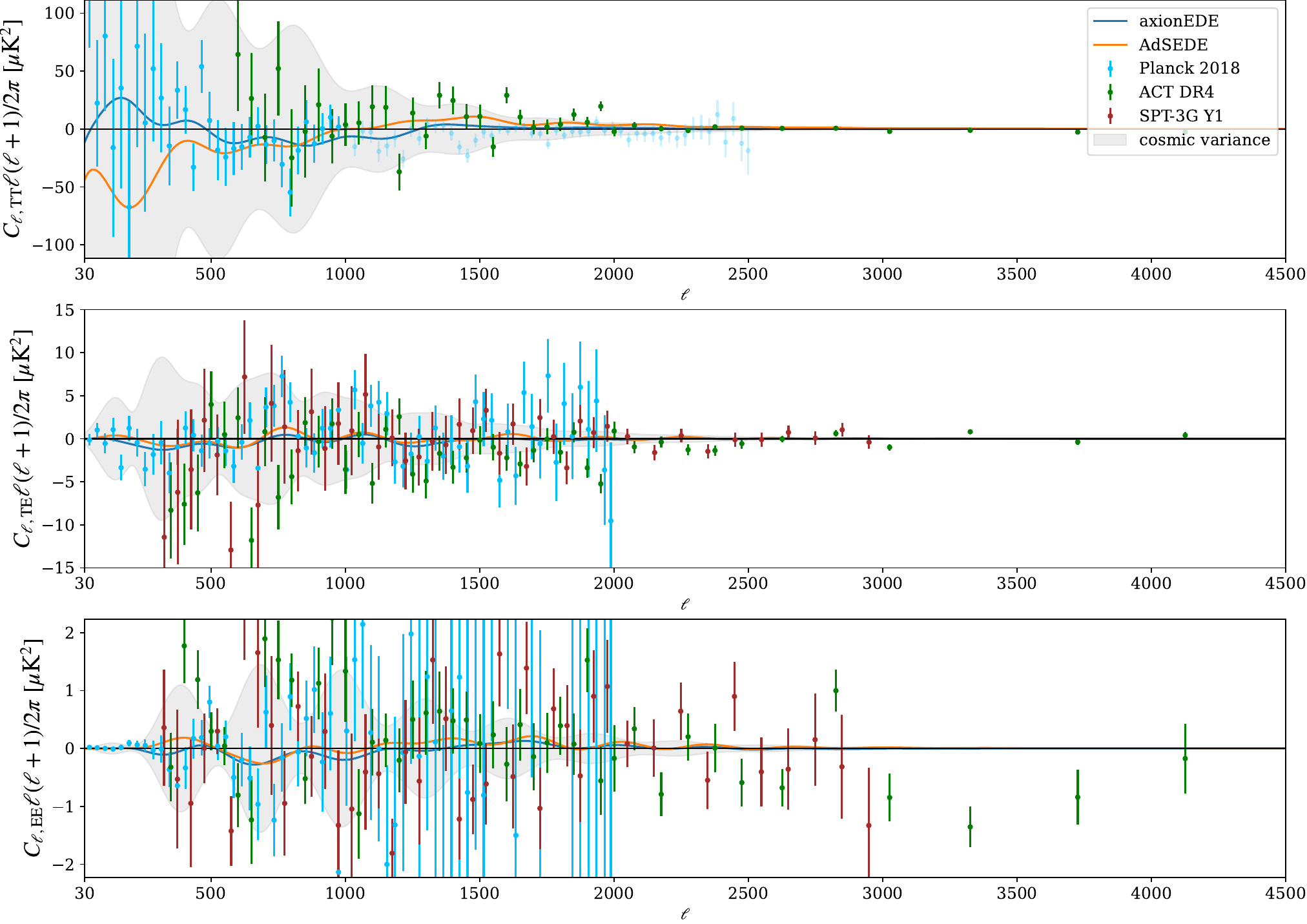}
\caption{\label{fig:residuals} The residuals of the best-fit
points for each model relative to $\Lambda$CDM model in Planck
($\ell _{\text{TT, max}} =1000$) + ACT DR4 + SPT-3G Y1 + BAO +
Pantheon and the constraints of different CMB data. The light
colored part of the Planck TT spectrum is the unused data.}
\end{figure}

\subsection{Where do the differences in model fit come from?}

As the main improvements come from the ACT and Planck high-$\ell$
part, we display the cumulative $\Delta \chi^2$ of them in
\autoref{fig:chi2} to clarify the origin of the data in favor of
the axion-like EDE model. It is clear that the preference for the
axion-like EDE arises mainly from the narrow $\ell \approx 1300$
part of the ACT TE spectrum and the wide $1400\lesssim \ell
\lesssim 2000$ part of the ACT TT spectrum, where the latter is
also valid for AdS-EDE. The lowest several bins of the ACT EE
spectrum also contribute to the improvement for both model
(similar to the analysis in \cite{Hill:2021yec}). However, the
next few bins could not be fitted well, resulting in no
significant preference in the ACT EE spectrum.

The exact reasons can be found in the residual plots
(\autoref{fig:residuals}). ACT data in the $\ell \approx 1300$
part of the TE spectrum seems to favor smaller values, and the
axion-like EDE captures this point. Besides, in the axion-like EDE
model, there seems to be a shift to larger $\ell$ since the 5th
acoustic peak of the TT spectrum. This shift is more visible in
the EE spectrum.

In addition, the axion-like EDE model also provides a better fit
to the $\ell \lesssim 900$ part of Planck TE,EE spectrum. Some of
them come from the previously mentioned shift toward high $\ell$.

\subsection{Why does the high-$\ell$ part of Planck TT not favor the EDE model?}

\begin{figure}[tbp]
\centering
\includegraphics[width=0.3\textwidth]{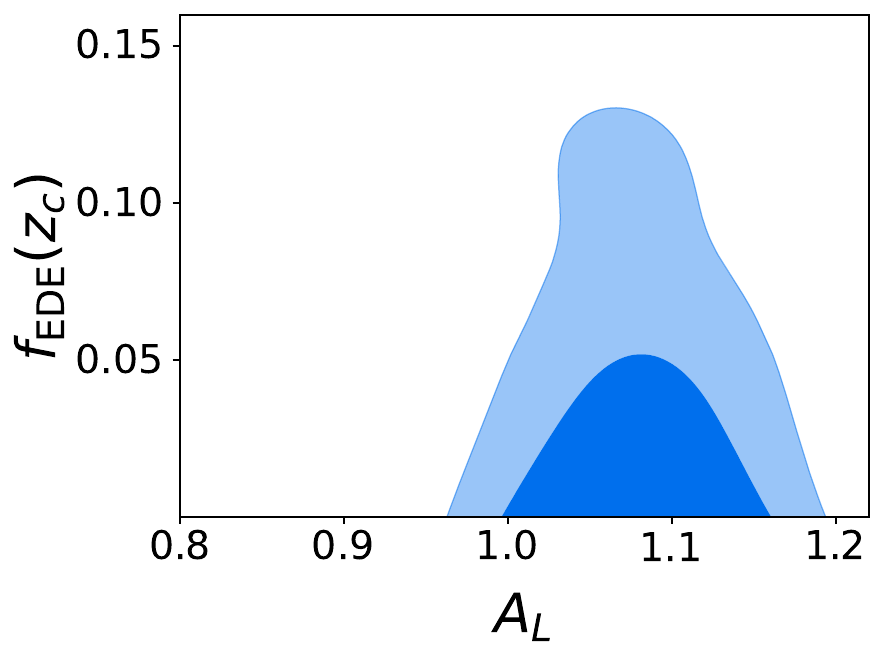}
\caption{\label{fig:AL} The relation between $A_L$ and
$f_\text{EDE}$ in axion-like EDE model with fullPlanck + BAO +
Pantheon dataset. }
\end{figure}

There are some oscillatory residuals in the Planck TT high-$\ell$
part, and such oscillatory residuals might be related to the
lensing anomaly \cite{Addison:2015wyg,Motloch:2019gux} and
parameter differences between different scales of the Planck TT
spectrum \cite{Addison:2015wyg,Planck:2016tof}. Although
oscillatory residuals also appear in our bestfit EDE model, it is
in a different phase than the oscillatory residuals of the
high-$\ell$ part of Planck TT spectrum, see
\autoref{fig:residuals}.

Nevertheless, we considered the simplest possibility: whether the
possible solution to the lensing anomaly is related to that the
high-$\ell$ part of Planck TT spectrum disfavors the EDE models?
We performed the MCMC analysis by varying the rescaling factor
$A_L$ (see \cite{Calabrese:2008rt} for standard definition) of the
lensing potential with the fullPlanck+BAO+Pantheon dataset. The
results are presented in \autoref{fig:AL}, where we do not find
any significant degeneracy between $A_L$ and EDE parameters, which
suggests that the lensing anomaly alone cannot explain this
disfavor to EDE models.
See Fig.6 in \cite{Wang:2022jpo} for an attempt to modify late dark energy.

However, it is important to note that the lensing anomaly is only
a naive reflection of the oscillation pattern on this Planck TT
spectrum. In fact, lensing anomaly also cannot explain the
inconsistency between different scales of the Planck TT spectrum
\cite{Planck:2016tof}. And this oscillation pattern was not found
in ground-based CMB observations (see also \cite{Lin:2020jcb}),
which can already be better than Planck constraints in $\ell
\gtrsim 2000$, so further observations of the TT spectrum in the
$1000 \lesssim \ell \lesssim 2000$ part are necessary.

\subsection{$n_s=1$?}

\begin{figure}[tbp]
\centering
\includegraphics[width=\textwidth]{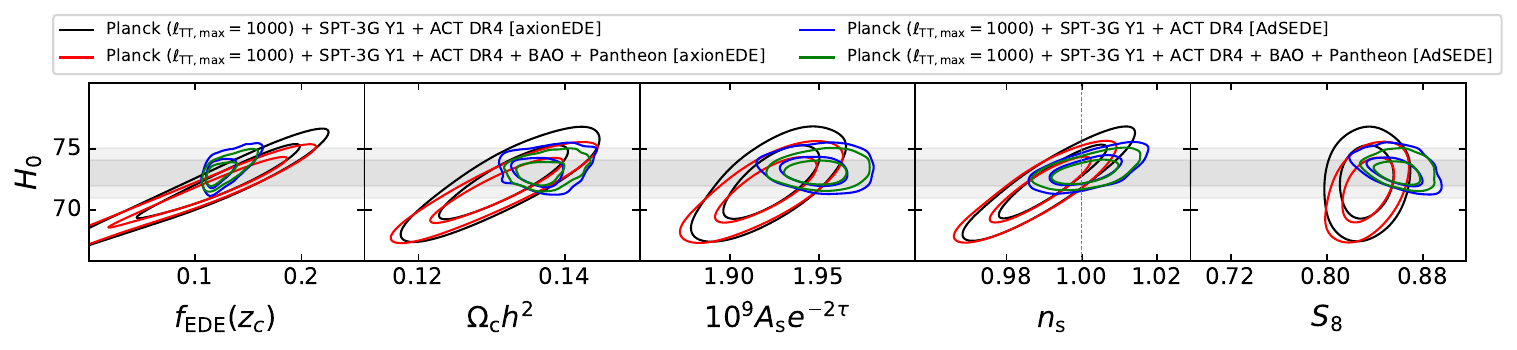}
\caption{\label{fig:parameters_shift} The degeneracy between
parameters under different models and datasets.}
\end{figure}

\begin{figure}[tbp]
\centering
\includegraphics[width=\textwidth]{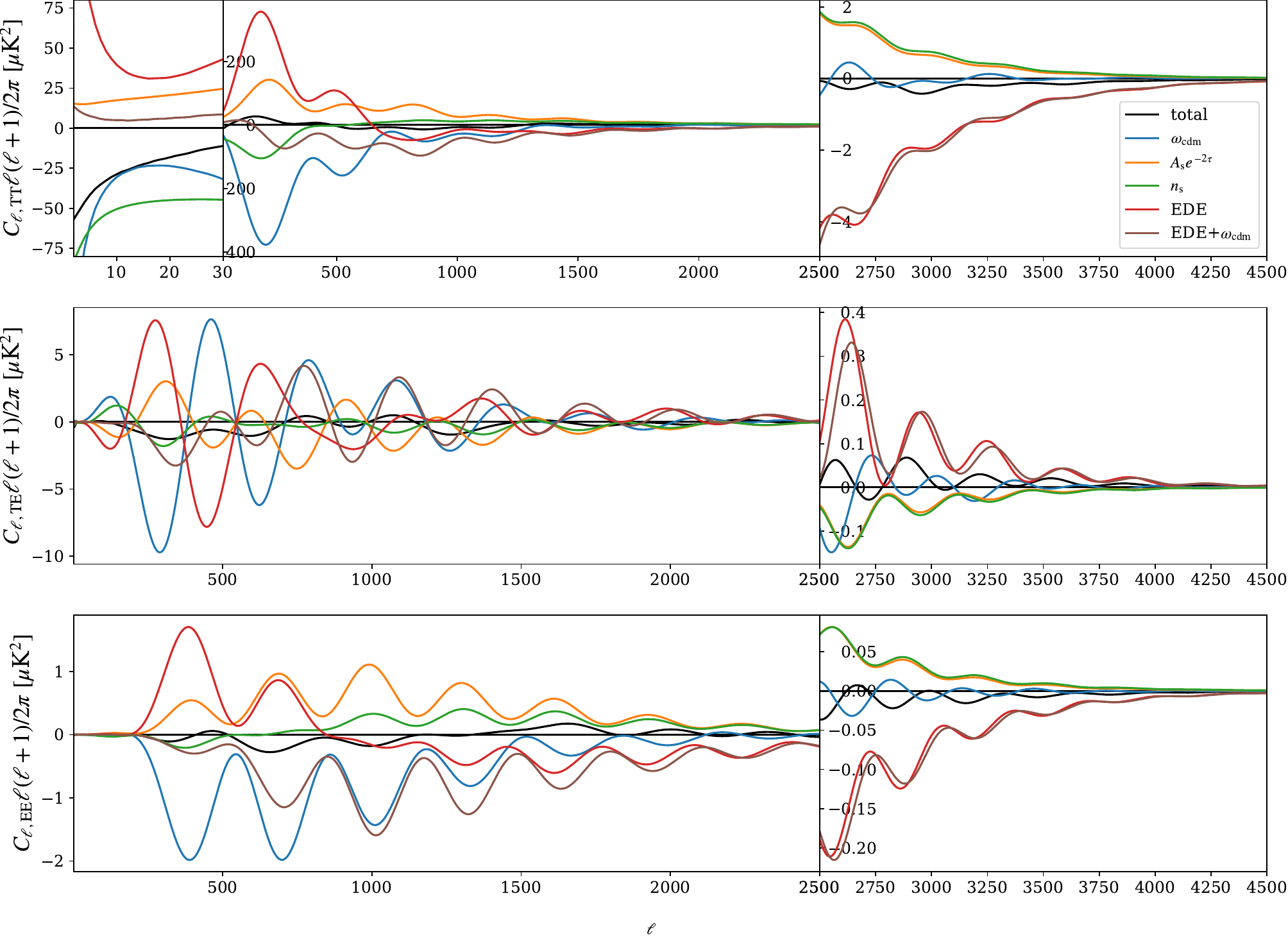}
\caption{\label{fig:specturm_parameter} Relative changes in CMB
power spectrum with respect to our bestfit $\Lambda$CDM model when
changing certain parameters from $\Lambda$CDM to axion-like EDE
bestfit values. $\theta_\text{s}^*$ is fixed by adjusting $H_0$
when changing $\omega_m$ and relevant EDE parameters.}
\end{figure}

\begin{figure}[tbp]
\centering
\includegraphics[width=0.9\textwidth]{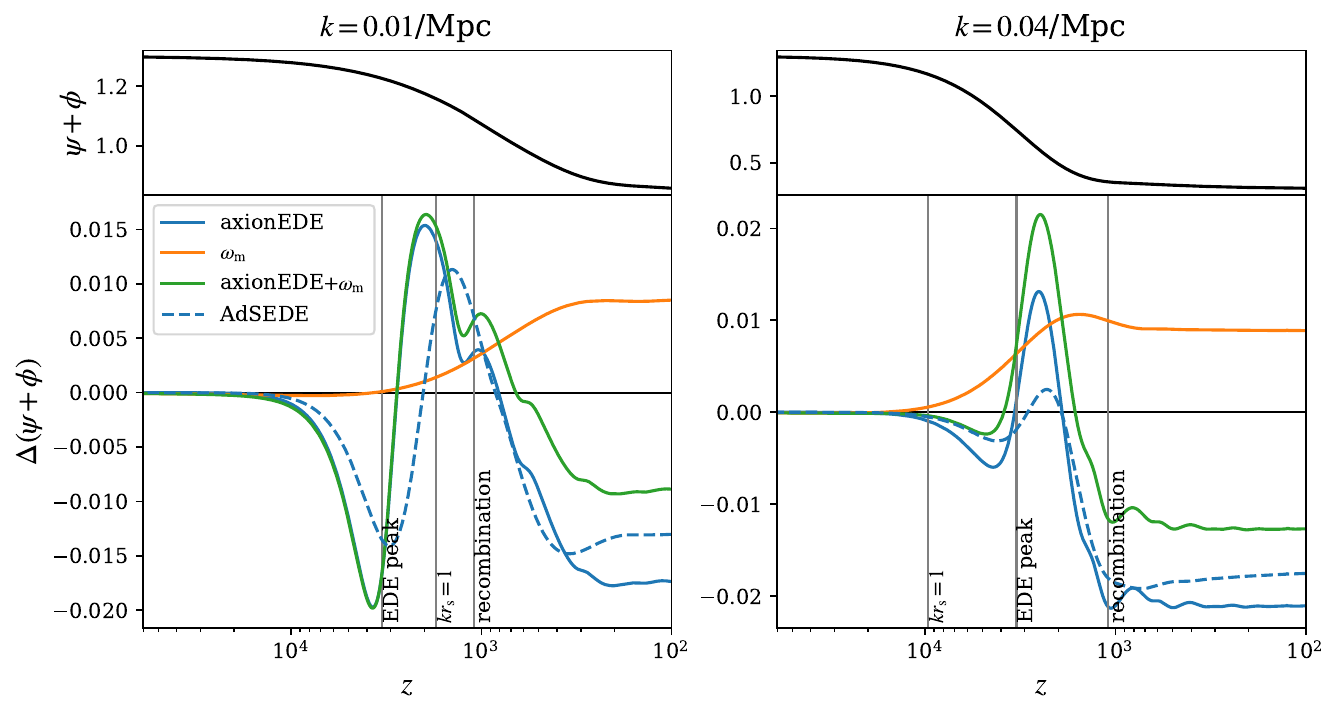}
\caption{\label{fig:two_k} Relative changes in Weyl potential with
respect to our bestfit $\Lambda$CDM model when changing certain
parameters from $\Lambda$CDM to EDE bestfit values.}
\end{figure}

In EDE models, some other cosmological parameters must be shifted
to compensate the raising of $H_0$. We show clear shifts of
relevant parameters in \autoref{fig:parameters_shift}. As these
parameters are shifted, the CMB power spectrum barely changes over
the range of scales we are interested in, as showed by the black
line in the \autoref{fig:specturm_parameter}. This indicates that
our CMB dataset is well constrained to the power spectrum.
\footnote{Note that some of the parameters are shifted differently than in Ref.\cite{Jiang:2021bab}. This is partly because the $\theta_s$ measurement is not as precise in Ref.\cite{Jiang:2021bab} as it is here.}

In addition to altering $r_\text{s}^*$, pre-recombination EDE also
affects the amplitude of power spectrum. Here, the Weyl potential
will accelerated decay (eventually to a smaller level), as shown
in \autoref{fig:two_k}, see also
\cite{Lin:2019qug,Niedermann:2020dwg} for other EDE models. The
faster decay of the Weyl potential results in the enhancement of
CMB power spectrum of $\ell \lesssim 700$, which further leads to
the enhancement of early ISW effect. In order to balance the
enhancement of the early ISW effect, which is measured to be
self-consistent in the $\Lambda$CDM model \cite{Vagnozzi:2021gjh},
$\omega_\text{cdm}$ is shifted to a larger value. The direction of
shift is consistent with the constraint of BAO+Pantheon. The
remaining power spectrum amplitude changes, especially around the
pivot, is complemented by the overall amplitude $A_\mathrm{s}
e^{-2\tau}$.

The high $\ell$ part of power spectrum is mainly controlled by
diffusion damping. The energy injection before recombination will
amplify the ratio $r_\text{d}/r_\text{s}^*$, thus enhancing the
damping as $\theta_\text{s}$ is well fixed. This can not be
complemented fully by the rise of $A_\mathrm{s} e^{-2\tau}$, leads
to the rise of $n_s$, as shown in \autoref{fig:parameters_shift}.
\footnote{The rises of $n_s$ and $\omega_\text{m}$ result in a
larger $S_8$, which, however, can be lowered by new physics beyond
cold dark matter \cite{Ye:2021iwa,Allali:2021azp,Clark:2021hlo,Luu:2021yhl,Haridasu:2020xaa}, see also 
\cite{DiValentino:2019ffd,Liu:2021mkv,Heisenberg:2022lob}.}

Actually, we have \begin{equation} \delta n_s \approx 0.3
\frac{\delta H_0}{H_0} \label{ns}\end{equation} for our combined
CMB dataset, which is slightly different for other dataset, e.g.
$\delta n_s \approx 0.4 \frac{\delta H_0}{H_0}$ for
fullPlanck+BAO+Pantheon dataset \cite{Ye:2021nej}, see also recent
\cite{Ye:2022afu}. This is due to
the different constraining power of the CMB datasets, especially
in the region of diffusion damping. However, they all suggest $n_s
= 1$, i.e. a scale invariant spectrum.
See also Refs.\cite{Benetti:2013wla,Benetti:2017gvm,Benetti:2017juy} for studies with $N_\text{eff}$.
This hints that the
primordial Universe model might need to be reconsidered, e.g.
a inflation potential $V \sim \phi^p$ with a small $p$ \cite{DAmico:2021zdd},
the curvaton scenario with a sub-Planckian field excursion \cite{Takahashi:2021bti},
multi-natural inflation \cite{Czerny:2014wza,Czerny:2014qqa,Croon:2014dma,Higaki:2015kta},
ALP or QCD axion inflation \cite{Daido:2017wwb,Daido:2017tbr,Takahashi:2019qmh,Takahashi:2019pqf,Takahashi:2021tff},
see also \cite{Takahashi:2013cxa}.

\section{Conclusion}
\label{sec:5}

We investigated the constraints on EDE by combining the most
precise CMB data available, i.e. Planck, SPT and ACT data.
However, since the small scale part of Planck TT spectrum suffers
from some anomalies, we exclude $\ell>1000$ part of Planck TT
power spectrum. Our main conclusions are as follows:
\begin{itemize}
    \item The combined CMB dataset favors non-zero EDE fractions and
large Hubble constants for both axion-like EDE and
    AdS-EDE models, i.e. $H_0=72.4(73.4)^{+2.2}_{-1.7}$ km/s/Mpc and
$73.31(72.98)^{+0.68}_{-0.93}$ km/s/Mpc, respectively. The
inclusion of BAO+Pantheon data does not change the results, i.e.
$H_0=71.6( 72.9)_{-1.5}^{+2.0}$ and
$73.17(72.74)^{+0.55}_{-0.77}$ km/s/Mpc, respectively. 
In addition, it is also noted that 
the fullPlanck+BAO+Pantheon dataset still favors a large $H_0 =
72.52(72.46)\pm0.51$ km/s/Mpc for AdS-EDE.
\item The axion-like
EDE model can fit the data significantly better
    than $\Lambda$CDM, which is mainly driven by the ACT part.
    \item The reason for the disfavor of the high-$\ell$ part of Planck TT to
    EDE is that the oscillatory pattern of high-$\ell$ TT spectrum in EDE do not
    match the oscillatory residual of
    Planck. The lensing anomaly can not explain it.
    \item The scale relationship between $n_s$ and $H_0$ reappears, see (\autoref{ns}), which implies $n_s=1$ for $H_0\sim
73$ km/s/Mpc.
\end{itemize}

However, it should be noted that the ACT data seem to have some slight discrepancies with Planck and SPT results \cite{ACT:2020gnv,Handley:2020hdp}
, at least under the $\Lambda$CDM model.
Therefore further observations and analyses on CMB are needed.
In fact, the first year result of SPT-3G only made use of half of
a observing season and part of detectors \cite{SPT-3G:2021eoc},
and also the results of ACT DR5 is being released
\cite{Mallaby-Kay:2021tuk}. Ground-based CMB observations will
have even better constraints, which will help to reveal whether
and how the CMB observations really favor the EDE models.


\textbf{Note added:} When this project will be completed,
Ref.\cite{Smith:2022hwi} is
present, which has investigated the constraints on (axion-like)
EDE using ACT DR4, SPT-3G, and Planck2018 ($\ell_\mathrm{TT}<650$) data,
with and without BAO+Pantheon data.

\acknowledgments

We thank Gen Ye for helpful discussions. This work is supported by
the NSFC, No.12075246, the KRPCAS, No.XDPB15. We acknowledge the
Tianhe-2 supercomputer for providing computing resources.

\bibliographystyle{JHEP}
\bibliography{./refs.bib}

\providecommand{\href}[2]{#2}\begingroup\raggedright\begin{thebibliography}{10}

\bibitem{Verde:2019ivm}
L.~Verde, T.~Treu, and A.~G. Riess, {\it {Tensions between the Early and the
  Late Universe}},  {\em Nature Astron.} {\bf 3} (7, 2019) 891,
  [\href{http://arxiv.org/abs/1907.10625}{{\tt arXiv:1907.10625}}].

\bibitem{Riess:2019qba}
A.~G. Riess, {\it {The Expansion of the Universe is Faster than Expected}},
  {\em Nature Rev. Phys.} {\bf 2} (2019), no.~1 10--12,
  [\href{http://arxiv.org/abs/2001.03624}{{\tt arXiv:2001.03624}}].

\bibitem{DiValentino:2020zio}
E.~Di~Valentino et~al., {\it {Snowmass2021 - Letter of interest cosmology
  intertwined II: The hubble constant tension}},  {\em Astropart. Phys.} {\bf
  131} (2021) 102605, [\href{http://arxiv.org/abs/2008.11284}{{\tt
  arXiv:2008.11284}}].

\bibitem{DiValentino:2021izs}
E.~Di~Valentino, O.~Mena, S.~Pan, L.~Visinelli, W.~Yang, A.~Melchiorri, D.~F.
  Mota, A.~G. Riess, and J.~Silk, {\it {In the realm of the Hubble
  tension\textemdash{}a review of solutions}},  {\em Class. Quant. Grav.} {\bf
  38} (2021), no.~15 153001, [\href{http://arxiv.org/abs/2103.01183}{{\tt
  arXiv:2103.01183}}].

\bibitem{Perivolaropoulos:2021jda}
L.~Perivolaropoulos and F.~Skara, {\it {Challenges for $\Lambda$CDM: An
  update}},  \href{http://arxiv.org/abs/2105.05208}{{\tt arXiv:2105.05208}}.

\bibitem{DiValentino:2016hlg}
E.~Di~Valentino, A.~Melchiorri, and J.~Silk, {\it {Reconciling Planck with the
  local value of $H_0$ in extended parameter space}},  {\em Phys. Lett. B} {\bf
  761} (2016) 242--246, [\href{http://arxiv.org/abs/1606.00634}{{\tt
  arXiv:1606.00634}}].

\bibitem{Mortsell:2018mfj}
E.~M\"ortsell and S.~Dhawan, {\it {Does the Hubble constant tension call for
  new physics?}},  {\em JCAP} {\bf 09} (2018) 025,
  [\href{http://arxiv.org/abs/1801.07260}{{\tt arXiv:1801.07260}}].

\bibitem{Vagnozzi:2019ezj}
S.~Vagnozzi, {\it {New physics in light of the $H_0$ tension: An alternative
  view}},  {\em Phys. Rev. D} {\bf 102} (2020), no.~2 023518,
  [\href{http://arxiv.org/abs/1907.07569}{{\tt arXiv:1907.07569}}].

\bibitem{Knox:2019rjx}
L.~Knox and M.~Millea, {\it {Hubble constant hunter\textquoteright{}s guide}},
  {\em Phys. Rev. D} {\bf 101} (2020), no.~4 043533,
  [\href{http://arxiv.org/abs/1908.03663}{{\tt arXiv:1908.03663}}].

\bibitem{Schoneberg:2021qvd}
N.~Sch\"oneberg, G.~Franco~Abell\'an, A.~P\'erez~S\'anchez, S.~J. Witte,
  V.~Poulin, and J.~Lesgourgues, {\it {The $H_0$ Olympics: A fair ranking of
  proposed models}},  \href{http://arxiv.org/abs/2107.10291}{{\tt
  arXiv:2107.10291}}.

\bibitem{Karwal:2016vyq}
T.~Karwal and M.~Kamionkowski, {\it {Dark energy at early times, the Hubble
  parameter, and the string axiverse}},  {\em Phys. Rev. D} {\bf 94} (2016),
  no.~10 103523, [\href{http://arxiv.org/abs/1608.01309}{{\tt
  arXiv:1608.01309}}].

\bibitem{Poulin:2018cxd}
V.~Poulin, T.~L. Smith, T.~Karwal, and M.~Kamionkowski, {\it {Early Dark Energy
  Can Resolve The Hubble Tension}},  {\em Phys. Rev. Lett.} {\bf 122} (2019),
  no.~22 221301, [\href{http://arxiv.org/abs/1811.04083}{{\tt
  arXiv:1811.04083}}].

\bibitem{Kaloper:2019lpl}
N.~Kaloper, {\it {Dark energy, $H_0$ and weak gravity conjecture}},  {\em Int.
  J. Mod. Phys. D} {\bf 28} (2019), no.~14 1944017,
  [\href{http://arxiv.org/abs/1903.11676}{{\tt arXiv:1903.11676}}].

\bibitem{Agrawal:2019lmo}
P.~Agrawal, F.-Y. Cyr-Racine, D.~Pinner, and L.~Randall, {\it {Rock 'n' Roll
  Solutions to the Hubble Tension}},
  \href{http://arxiv.org/abs/1904.01016}{{\tt arXiv:1904.01016}}.

\bibitem{Lin:2019qug}
M.-X. Lin, G.~Benevento, W.~Hu, and M.~Raveri, {\it {Acoustic Dark Energy:
  Potential Conversion of the Hubble Tension}},  {\em Phys. Rev. D} {\bf 100}
  (2019), no.~6 063542, [\href{http://arxiv.org/abs/1905.12618}{{\tt
  arXiv:1905.12618}}].

\bibitem{Smith:2019ihp}
T.~L. Smith, V.~Poulin, and M.~A. Amin, {\it {Oscillating scalar fields and the
  Hubble tension: a resolution with novel signatures}},  {\em Phys. Rev. D}
  {\bf 101} (2020), no.~6 063523, [\href{http://arxiv.org/abs/1908.06995}{{\tt
  arXiv:1908.06995}}].

\bibitem{Niedermann:2019olb}
F.~Niedermann and M.~S. Sloth, {\it {New early dark energy}},  {\em Phys. Rev.
  D} {\bf 103} (2021), no.~4 L041303,
  [\href{http://arxiv.org/abs/1910.10739}{{\tt arXiv:1910.10739}}].

\bibitem{Sakstein:2019fmf}
J.~Sakstein and M.~Trodden, {\it {Early Dark Energy from Massive Neutrinos as a
  Natural Resolution of the Hubble Tension}},  {\em Phys. Rev. Lett.} {\bf 124}
  (2020), no.~16 161301, [\href{http://arxiv.org/abs/1911.11760}{{\tt
  arXiv:1911.11760}}].

\bibitem{Ye:2020btb}
G.~Ye and Y.-S. Piao, {\it {Is the Hubble tension a hint of AdS phase around
  recombination?}},  {\em Phys. Rev. D} {\bf 101} (2020), no.~8 083507,
  [\href{http://arxiv.org/abs/2001.02451}{{\tt arXiv:2001.02451}}].

\bibitem{Gogoi:2020qif}
A.~Gogoi, R.~K. Sharma, P.~Chanda, and S.~Das, {\it {Early Mass-varying
  Neutrino Dark Energy: Nugget Formation and Hubble Anomaly}},  {\em Astrophys.
  J.} {\bf 915} (2021), no.~2 132, [\href{http://arxiv.org/abs/2005.11889}{{\tt
  arXiv:2005.11889}}].

\bibitem{Braglia:2020bym}
M.~Braglia, W.~T. Emond, F.~Finelli, A.~E. Gumrukcuoglu, and K.~Koyama, {\it
  {Unified framework for early dark energy from $\alpha$-attractors}},  {\em
  Phys. Rev. D} {\bf 102} (2020), no.~8 083513,
  [\href{http://arxiv.org/abs/2005.14053}{{\tt arXiv:2005.14053}}].

\bibitem{Lin:2020jcb}
M.-X. Lin, W.~Hu, and M.~Raveri, {\it {Testing $H_0$ in Acoustic Dark Energy
  with Planck and ACT Polarization}},  {\em Phys. Rev. D} {\bf 102} (2020)
  123523, [\href{http://arxiv.org/abs/2009.08974}{{\tt arXiv:2009.08974}}].

\bibitem{Seto:2021xua}
O.~Seto and Y.~Toda, {\it {Comparing early dark energy and extra radiation
  solutions to the Hubble tension with BBN}},  {\em Phys. Rev. D} {\bf 103}
  (2021), no.~12 123501, [\href{http://arxiv.org/abs/2101.03740}{{\tt
  arXiv:2101.03740}}].

\bibitem{Nojiri:2021dze}
S.~Nojiri, S.~D. Odintsov, D.~Saez-Chillon~Gomez, and G.~S. Sharov, {\it
  {Modeling and testing the equation of state for (Early) dark energy}},  {\em
  Phys. Dark Univ.} {\bf 32} (2021) 100837,
  [\href{http://arxiv.org/abs/2103.05304}{{\tt arXiv:2103.05304}}].

\bibitem{Karwal:2021vpk}
T.~Karwal, M.~Raveri, B.~Jain, J.~Khoury, and M.~Trodden, {\it {Chameleon early
  dark energy and the Hubble tension}},  {\em Phys. Rev. D} {\bf 105} (2022),
  no.~6 063535, [\href{http://arxiv.org/abs/2106.13290}{{\tt
  arXiv:2106.13290}}].

\bibitem{Zumalacarregui:2020cjh}
M.~Zumalacarregui, {\it {Gravity in the Era of Equality: Towards solutions to
  the Hubble problem without fine-tuned initial conditions}},  {\em Phys. Rev.
  D} {\bf 102} (2020), no.~2 023523,
  [\href{http://arxiv.org/abs/2003.06396}{{\tt arXiv:2003.06396}}].

\bibitem{Ballesteros:2020sik}
G.~Ballesteros, A.~Notari, and F.~Rompineve, {\it {The $H_0$ tension: $\Delta
  G_N$ vs. $\Delta N_{\rm eff}$}},  {\em JCAP} {\bf 11} (2020) 024,
  [\href{http://arxiv.org/abs/2004.05049}{{\tt arXiv:2004.05049}}].

\bibitem{Braglia:2020auw}
M.~Braglia, M.~Ballardini, F.~Finelli, and K.~Koyama, {\it {Early modified
  gravity in light of the $H_0$ tension and LSS data}},  {\em Phys. Rev. D}
  {\bf 103} (2021), no.~4 043528, [\href{http://arxiv.org/abs/2011.12934}{{\tt
  arXiv:2011.12934}}].

\bibitem{Hill:2020osr}
J.~C. Hill, E.~McDonough, M.~W. Toomey, and S.~Alexander, {\it {Early dark
  energy does not restore cosmological concordance}},  {\em Phys. Rev. D} {\bf
  102} (2020), no.~4 043507, [\href{http://arxiv.org/abs/2003.07355}{{\tt
  arXiv:2003.07355}}].

\bibitem{Addison:2015wyg}
G.~E. Addison, Y.~Huang, D.~J. Watts, C.~L. Bennett, M.~Halpern, G.~Hinshaw,
  and J.~L. Weiland, {\it {Quantifying discordance in the 2015 Planck CMB
  spectrum}},  {\em Astrophys. J.} {\bf 818} (2016), no.~2 132,
  [\href{http://arxiv.org/abs/1511.00055}{{\tt arXiv:1511.00055}}].

\bibitem{Planck:2016tof}
{\bf Planck} Collaboration, N.~Aghanim et~al., {\it {Planck intermediate
  results. LI. Features in the cosmic microwave background temperature power
  spectrum and shifts in cosmological parameters}},  {\em Astron. Astrophys.}
  {\bf 607} (2017) A95, [\href{http://arxiv.org/abs/1608.02487}{{\tt
  arXiv:1608.02487}}].

\bibitem{Motloch:2019gux}
P.~Motloch and W.~Hu, {\it {Lensinglike tensions in the $Planck$ legacy
  release}},  {\em Phys. Rev. D} {\bf 101} (2020), no.~8 083515,
  [\href{http://arxiv.org/abs/1912.06601}{{\tt arXiv:1912.06601}}].

\bibitem{SPT:2017jdf}
{\bf SPT} Collaboration, J.~W. Henning et~al., {\it {Measurements of the
  Temperature and E-Mode Polarization of the CMB from 500 Square Degrees of
  SPTpol Data}},  {\em Astrophys. J.} {\bf 852} (2018), no.~2 97,
  [\href{http://arxiv.org/abs/1707.09353}{{\tt arXiv:1707.09353}}].

\bibitem{ACT:2020gnv}
{\bf ACT} Collaboration, S.~Aiola et~al., {\it {The Atacama Cosmology
  Telescope: DR4 Maps and Cosmological Parameters}},  {\em JCAP} {\bf 12}
  (2020) 047, [\href{http://arxiv.org/abs/2007.07288}{{\tt arXiv:2007.07288}}].

\bibitem{SPT-3G:2021eoc}
{\bf SPT-3G} Collaboration, D.~Dutcher et~al., {\it {Measurements of the E-mode
  polarization and temperature-E-mode correlation of the CMB from SPT-3G 2018
  data}},  {\em Phys. Rev. D} {\bf 104} (2021), no.~2 022003,
  [\href{http://arxiv.org/abs/2101.01684}{{\tt arXiv:2101.01684}}].

\bibitem{Chudaykin:2020acu}
A.~Chudaykin, D.~Gorbunov, and N.~Nedelko, {\it {Combined analysis of Planck
  and SPTPol data favors the early dark energy models}},  {\em JCAP} {\bf 08}
  (2020) 013, [\href{http://arxiv.org/abs/2004.13046}{{\tt arXiv:2004.13046}}].

\bibitem{Chudaykin:2020igl}
A.~Chudaykin, D.~Gorbunov, and N.~Nedelko, {\it {Exploring an early dark energy
  solution to the Hubble tension with Planck and SPTPol data}},  {\em Phys.
  Rev. D} {\bf 103} (2021), no.~4 043529,
  [\href{http://arxiv.org/abs/2011.04682}{{\tt arXiv:2011.04682}}].

\bibitem{Jiang:2021bab}
J.-Q. Jiang and Y.-S. Piao, {\it {Testing AdS early dark energy with Planck,
  SPTpol, and LSS data}},  {\em Phys. Rev. D} {\bf 104} (2021), no.~10 103524,
  [\href{http://arxiv.org/abs/2107.07128}{{\tt arXiv:2107.07128}}].

\bibitem{Hill:2021yec}
J.~C. Hill et~al., {\it {The Atacama Cosmology Telescope: Constraints on
  Pre-Recombination Early Dark Energy}},
  \href{http://arxiv.org/abs/2109.04451}{{\tt arXiv:2109.04451}}.

\bibitem{Poulin:2021bjr}
V.~Poulin, T.~L. Smith, and A.~Bartlett, {\it {Dark energy at early times and
  ACT data: A larger Hubble constant without late-time priors}},  {\em Phys.
  Rev. D} {\bf 104} (2021), no.~12 123550,
  [\href{http://arxiv.org/abs/2109.06229}{{\tt arXiv:2109.06229}}].

\bibitem{LaPosta:2021pgm}
A.~La~Posta, T.~Louis, X.~Garrido, and J.~C. Hill, {\it {Constraints on
  Pre-Recombination Early Dark Energy from SPT-3G Public Data}},
  \href{http://arxiv.org/abs/2112.10754}{{\tt arXiv:2112.10754}}.

\bibitem{Ye:2021nej}
G.~Ye, B.~Hu, and Y.-S. Piao, {\it {Implication of the Hubble tension for the
  primordial Universe in light of recent cosmological data}},  {\em Phys. Rev.
  D} {\bf 104} (2021), no.~6 063510,
  [\href{http://arxiv.org/abs/2103.09729}{{\tt arXiv:2103.09729}}].

\bibitem{Ye:2020oix}
G.~Ye and Y.-S. Piao, {\it {$T_0$ censorship of early dark energy and AdS
  vacua}},  {\em Phys. Rev. D} {\bf 102} (2020), no.~8 083523,
  [\href{http://arxiv.org/abs/2008.10832}{{\tt arXiv:2008.10832}}].

\bibitem{Akarsu:2019hmw}
O.~Akarsu, J.~D. Barrow, L.~A. Escamilla, and J.~A. Vazquez, {\it {Graduated
  dark energy: Observational hints of a spontaneous sign switch in the
  cosmological constant}},  {\em Phys. Rev. D} {\bf 101} (2020), no.~6 063528,
  [\href{http://arxiv.org/abs/1912.08751}{{\tt arXiv:1912.08751}}].

\bibitem{Visinelli:2019qqu}
L.~Visinelli, S.~Vagnozzi, and U.~Danielsson, {\it {Revisiting a negative
  cosmological constant from low-redshift data}},  {\em Symmetry} {\bf 11}
  (2019), no.~8 1035, [\href{http://arxiv.org/abs/1907.07953}{{\tt
  arXiv:1907.07953}}].

\bibitem{Dutta:2018vmq}
K.~Dutta, Ruchika, A.~Roy, A.~A. Sen, and M.~M. Sheikh-Jabbari, {\it {Beyond
  $\Lambda $CDM with low and high redshift data: implications for dark
  energy}},  {\em Gen. Rel. Grav.} {\bf 52} (2020), no.~2 15,
  [\href{http://arxiv.org/abs/1808.06623}{{\tt arXiv:1808.06623}}].

\bibitem{Calderon:2020hoc}
R.~Calder\'on, R.~Gannouji, B.~L'Huillier, and D.~Polarski, {\it {Negative
  cosmological constant in the dark sector?}},  {\em Phys. Rev. D} {\bf 103}
  (2021), no.~2 023526, [\href{http://arxiv.org/abs/2008.10237}{{\tt
  arXiv:2008.10237}}].

\bibitem{Ruchika:2020avj}
Ruchika, K.~Dutta, A.~Mukherjee, and A.~A. Sen, {\it {Observational Constraints
  on Axion(s) with a Cosmological Constant}},
  \href{http://arxiv.org/abs/2005.08813}{{\tt arXiv:2005.08813}}.

\bibitem{Akarsu:2021fol}
O.~Akarsu, S.~Kumar, E.~\"Oz\"ulker, and J.~A. Vazquez, {\it {Relaxing
  cosmological tensions with a sign switching cosmological constant}},  {\em
  Phys. Rev. D} {\bf 104} (2021), no.~12 123512,
  [\href{http://arxiv.org/abs/2108.09239}{{\tt arXiv:2108.09239}}].

\bibitem{Sen:2021wld}
A.~A. Sen, S.~A. Adil, and S.~Sen, {\it {Do cosmological observations allow a
  negative $\Lambda$?}},  \href{http://arxiv.org/abs/2112.10641}{{\tt
  arXiv:2112.10641}}.

\bibitem{Planck:2018vyg}
{\bf Planck} Collaboration, N.~Aghanim et~al., {\it {Planck 2018 results. VI.
  Cosmological parameters}},  {\em Astron. Astrophys.} {\bf 641} (2020) A6,
  [\href{http://arxiv.org/abs/1807.06209}{{\tt arXiv:1807.06209}}]. [Erratum:
  Astron.Astrophys. 652, C4 (2021)].

\bibitem{ACT:2020frw}
{\bf ACT} Collaboration, S.~K. Choi et~al., {\it {The Atacama Cosmology
  Telescope: a measurement of the Cosmic Microwave Background power spectra at
  98 and 150 GHz}},  {\em JCAP} {\bf 12} (2020) 045,
  [\href{http://arxiv.org/abs/2007.07289}{{\tt arXiv:2007.07289}}].

\bibitem{SPT-3G:2021wgf}
{\bf SPT-3G} Collaboration, L.~Balkenhol et~al., {\it {Constraints on
  \ensuremath{\Lambda}CDM extensions from the SPT-3G 2018 EE and TE power
  spectra}},  {\em Phys. Rev. D} {\bf 104} (2021), no.~8 083509,
  [\href{http://arxiv.org/abs/2103.13618}{{\tt arXiv:2103.13618}}].

\bibitem{Beutler:2011hx}
F.~Beutler, C.~Blake, M.~Colless, D.~H. Jones, L.~Staveley-Smith, L.~Campbell,
  Q.~Parker, W.~Saunders, and F.~Watson, {\it {The 6dF Galaxy Survey: Baryon
  Acoustic Oscillations and the Local Hubble Constant}},  {\em Mon. Not. Roy.
  Astron. Soc.} {\bf 416} (2011) 3017--3032,
  [\href{http://arxiv.org/abs/1106.3366}{{\tt arXiv:1106.3366}}].

\bibitem{Ross:2014qpa}
A.~J. Ross, L.~Samushia, C.~Howlett, W.~J. Percival, A.~Burden, and M.~Manera,
  {\it {The clustering of the SDSS DR7 main Galaxy sample \textendash{} I. A 4
  per cent distance measure at $z = 0.15$}},  {\em Mon. Not. Roy. Astron. Soc.}
  {\bf 449} (2015), no.~1 835--847, [\href{http://arxiv.org/abs/1409.3242}{{\tt
  arXiv:1409.3242}}].

\bibitem{BOSS:2016wmc}
{\bf BOSS} Collaboration, S.~Alam et~al., {\it {The clustering of galaxies in
  the completed SDSS-III Baryon Oscillation Spectroscopic Survey: cosmological
  analysis of the DR12 galaxy sample}},  {\em Mon. Not. Roy. Astron. Soc.} {\bf
  470} (2017), no.~3 2617--2652, [\href{http://arxiv.org/abs/1607.03155}{{\tt
  arXiv:1607.03155}}].

\bibitem{Scolnic:2017caz}
{\bf Pan-STARRS1} Collaboration, D.~M. Scolnic et~al., {\it {The Complete
  Light-curve Sample of Spectroscopically Confirmed SNe Ia from Pan-STARRS1 and
  Cosmological Constraints from the Combined Pantheon Sample}},  {\em
  Astrophys. J.} {\bf 859} (2018), no.~2 101,
  [\href{http://arxiv.org/abs/1710.00845}{{\tt arXiv:1710.00845}}].

\bibitem{Torrado:2020dgo}
J.~Torrado and A.~Lewis, {\it {Cobaya: Code for Bayesian Analysis of
  hierarchical physical models}},  {\em JCAP} {\bf 05} (2021) 057,
  [\href{http://arxiv.org/abs/2005.05290}{{\tt arXiv:2005.05290}}].

\bibitem{Blas:2011rf}
D.~Blas, J.~Lesgourgues, and T.~Tram, {\it {The Cosmic Linear Anisotropy
  Solving System (CLASS) II: Approximation schemes}},  {\em JCAP} {\bf 07}
  (2011) 034, [\href{http://arxiv.org/abs/1104.2933}{{\tt arXiv:1104.2933}}].

\bibitem{Lewis:2019xzd}
A.~Lewis, {\it {GetDist: a Python package for analysing Monte Carlo samples}},
  \href{http://arxiv.org/abs/1910.13970}{{\tt arXiv:1910.13970}}.

\bibitem{cartis2019improving}
C.~Cartis, J.~Fiala, B.~Marteau, and L.~Roberts, {\it Improving the flexibility
  and robustness of model-based derivative-free optimization solvers},  {\em
  ACM Transactions on Mathematical Software (TOMS)} {\bf 45} (2019), no.~3
  1--41.

\bibitem{cartis2018escaping}
C.~Cartis, L.~Roberts, and O.~Sheridan-Methven, {\it Escaping local minima with
  derivative-free methods: a numerical investigation},  {\em arXiv preprint
  arXiv:1812.11343} (2018).

\bibitem{powell2009bobyqa}
M.~J. Powell, {\it The bobyqa algorithm for bound constrained optimization
  without derivatives},  {\em Cambridge NA Report NA2009/06, University of
  Cambridge, Cambridge} {\bf 26} (2009).

\bibitem{Riess:2021jrx}
A.~G. Riess et~al., {\it {A Comprehensive Measurement of the Local Value of the
  Hubble Constant with 1 km/s/Mpc Uncertainty from the Hubble Space Telescope
  and the SH0ES Team}},  \href{http://arxiv.org/abs/2112.04510}{{\tt
  arXiv:2112.04510}}.

\bibitem{2112.10754}
A.~La~Posta, T.~Louis, X.~Garrido, and J.~C. Hill, {\it {Constraints on
  Pre-Recombination Early Dark Energy from SPT-3G Public Data}},
  \href{http://arxiv.org/abs/2112.10754}{{\tt arXiv:2112.10754}}.

\bibitem{Bernal:2016gxb}
J.~L. Bernal, L.~Verde, and A.~G. Riess, {\it {The trouble with $H_0$}},  {\em
  JCAP} {\bf 10} (2016) 019, [\href{http://arxiv.org/abs/1607.05617}{{\tt
  arXiv:1607.05617}}].

\bibitem{Aylor:2018drw}
K.~Aylor, M.~Joy, L.~Knox, M.~Millea, S.~Raghunathan, and W.~L.~K. Wu, {\it
  {Sounds Discordant: Classical Distance Ladder \& $\Lambda$CDM -based
  Determinations of the Cosmological Sound Horizon}},  {\em Astrophys. J.} {\bf
  874} (2019), no.~1 4, [\href{http://arxiv.org/abs/1811.00537}{{\tt
  arXiv:1811.00537}}].

\bibitem{Calabrese:2008rt}
E.~Calabrese, A.~Slosar, A.~Melchiorri, G.~F. Smoot, and O.~Zahn, {\it {Cosmic
  Microwave Weak lensing data as a test for the dark universe}},  {\em Phys.
  Rev. D} {\bf 77} (2008) 123531, [\href{http://arxiv.org/abs/0803.2309}{{\tt
  arXiv:0803.2309}}].

\bibitem{Wang:2022jpo}
H.~Wang and Y.-S. Piao, {\it {Testing dark energy after pre-recombination early
  dark energy}},  \href{http://arxiv.org/abs/2201.07079}{{\tt
  arXiv:2201.07079}}.

\bibitem{Niedermann:2020dwg}
F.~Niedermann and M.~S. Sloth, {\it {Resolving the Hubble tension with new
  early dark energy}},  {\em Phys. Rev. D} {\bf 102} (2020), no.~6 063527,
  [\href{http://arxiv.org/abs/2006.06686}{{\tt arXiv:2006.06686}}].

\bibitem{Vagnozzi:2021gjh}
S.~Vagnozzi, {\it {Consistency tests of \ensuremath{\Lambda}CDM from the early
  integrated Sachs-Wolfe effect: Implications for early-time new physics and
  the Hubble tension}},  {\em Phys. Rev. D} {\bf 104} (2021), no.~6 063524,
  [\href{http://arxiv.org/abs/2105.10425}{{\tt arXiv:2105.10425}}].

\bibitem{Ye:2021iwa}
G.~Ye, J.~Zhang, and Y.-S. Piao, {\it {Resolving both $H_0$ and $S_8$ tensions
  with AdS early dark energy and ultralight axion}},
  \href{http://arxiv.org/abs/2107.13391}{{\tt arXiv:2107.13391}}.

\bibitem{Allali:2021azp}
I.~J. Allali, M.~P. Hertzberg, and F.~Rompineve, {\it {Dark sector to restore
  cosmological concordance}},  {\em Phys. Rev. D} {\bf 104} (2021), no.~8
  L081303, [\href{http://arxiv.org/abs/2104.12798}{{\tt arXiv:2104.12798}}].

\bibitem{Clark:2021hlo}
S.~J. Clark, K.~Vattis, J.~Fan, and S.~M. Koushiappas, {\it {The $H_0$ and
  $S_8$ tensions necessitate early and late time changes to $\Lambda$CDM}},
  \href{http://arxiv.org/abs/2110.09562}{{\tt arXiv:2110.09562}}.

\bibitem{Luu:2021yhl}
H.~N. Luu, {\it {Axi-Higgs cosmology: Cosmic Microwave Background and
  cosmological tensions}},  \href{http://arxiv.org/abs/2111.01347}{{\tt
  arXiv:2111.01347}}.

\bibitem{Haridasu:2020xaa}
B.~S. Haridasu and M.~Viel, {\it {Late-time decaying dark matter: constraints
  and implications for the $H_0$-tension}},  {\em Mon. Not. Roy. Astron. Soc.}
  {\bf 497} (2020), no.~2 1757--1764,
  [\href{http://arxiv.org/abs/2004.07709}{{\tt arXiv:2004.07709}}].

\bibitem{DiValentino:2019ffd}
E.~Di~Valentino, A.~Melchiorri, O.~Mena, and S.~Vagnozzi, {\it {Interacting
  dark energy in the early 2020s: A promising solution to the $H_0$ and cosmic
  shear tensions}},  {\em Phys. Dark Univ.} {\bf 30} (2020) 100666,
  [\href{http://arxiv.org/abs/1908.04281}{{\tt arXiv:1908.04281}}].

\bibitem{Liu:2021mkv}
W.~Liu, L.~A. Anchordoqui, E.~Di~Valentino, S.~Pan, Y.~Wu, and W.~Yang, {\it
  {Constraints from high-precision measurements of the cosmic microwave
  background: the case of disintegrating dark matter with \ensuremath{\Lambda}
  or dynamical dark energy}},  {\em JCAP} {\bf 02} (2022), no.~02 012,
  [\href{http://arxiv.org/abs/2108.04188}{{\tt arXiv:2108.04188}}].

\bibitem{Heisenberg:2022lob}
L.~Heisenberg, H.~Villarrubia-Rojo, and J.~Zosso, {\it {Simultaneously solving
  the $H_0$ and $\sigma_8$ tensions with late dark energy}},
  \href{http://arxiv.org/abs/2201.11623}{{\tt arXiv:2201.11623}}.

\bibitem{Ye:2022afu}
G.~Ye and Y.-S. Piao, {\it {Improved constraint on primordial gravitational
  waves in light of the Hubble tension and BICEP/Keck}},
  \href{http://arxiv.org/abs/2202.10055}{{\tt arXiv:2202.10055}}.

\bibitem{Benetti:2013wla}
M.~Benetti, M.~Gerbino, W.~H. Kinney, E.~W. Kolb, M.~Lattanzi, A.~Melchiorri,
  L.~Pagano, and A.~Riotto, {\it {Cosmological data and indications for new
  physics}},  {\em JCAP} {\bf 10} (2013) 030,
  [\href{http://arxiv.org/abs/1303.4317}{{\tt arXiv:1303.4317}}].

\bibitem{Benetti:2017gvm}
M.~Benetti, L.~L. Graef, and J.~S. Alcaniz, {\it {Do joint CMB and HST data
  support a scale invariant spectrum?}},  {\em JCAP} {\bf 04} (2017) 003,
  [\href{http://arxiv.org/abs/1702.06509}{{\tt arXiv:1702.06509}}].

\bibitem{Benetti:2017juy}
M.~Benetti, L.~L. Graef, and J.~S. Alcaniz, {\it {The $H_0$ and $\sigma_8$
  tensions and the scale invariant spectrum}},  {\em JCAP} {\bf 07} (2018) 066,
  [\href{http://arxiv.org/abs/1712.00677}{{\tt arXiv:1712.00677}}].

\bibitem{DAmico:2021zdd}
G.~D'Amico, N.~Kaloper, and A.~Westphal, {\it {Very Hairy Inflation}},
  \href{http://arxiv.org/abs/2112.13861}{{\tt arXiv:2112.13861}}.

\bibitem{Takahashi:2021bti}
F.~Takahashi and W.~Yin, {\it {Cosmological implications of $n_s\approx 1$ in
  light of the Hubble tension}},  \href{http://arxiv.org/abs/2112.06710}{{\tt
  arXiv:2112.06710}}.

\bibitem{Czerny:2014wza}
M.~Czerny and F.~Takahashi, {\it {Multi-Natural Inflation}},  {\em Phys. Lett.
  B} {\bf 733} (2014) 241--246, [\href{http://arxiv.org/abs/1401.5212}{{\tt
  arXiv:1401.5212}}].

\bibitem{Czerny:2014qqa}
M.~Czerny, T.~Higaki, and F.~Takahashi, {\it {Multi-Natural Inflation in
  Supergravity and BICEP2}},  {\em Phys. Lett. B} {\bf 734} (2014) 167--172,
  [\href{http://arxiv.org/abs/1403.5883}{{\tt arXiv:1403.5883}}].

\bibitem{Croon:2014dma}
D.~Croon and V.~Sanz, {\it {Saving Natural Inflation}},  {\em JCAP} {\bf 02}
  (2015) 008, [\href{http://arxiv.org/abs/1411.7809}{{\tt arXiv:1411.7809}}].

\bibitem{Higaki:2015kta}
T.~Higaki and F.~Takahashi, {\it {Elliptic inflation: interpolating from
  natural inflation to R$^{2}$-inflation}},  {\em JHEP} {\bf 03} (2015) 129,
  [\href{http://arxiv.org/abs/1501.02354}{{\tt arXiv:1501.02354}}].

\bibitem{Daido:2017wwb}
R.~Daido, F.~Takahashi, and W.~Yin, {\it {The ALP miracle: unified inflaton and
  dark matter}},  {\em JCAP} {\bf 05} (2017) 044,
  [\href{http://arxiv.org/abs/1702.03284}{{\tt arXiv:1702.03284}}].

\bibitem{Daido:2017tbr}
R.~Daido, F.~Takahashi, and W.~Yin, {\it {The ALP miracle revisited}},  {\em
  JHEP} {\bf 02} (2018) 104, [\href{http://arxiv.org/abs/1710.11107}{{\tt
  arXiv:1710.11107}}].

\bibitem{Takahashi:2019qmh}
F.~Takahashi and W.~Yin, {\it {ALP inflation and Big Bang on Earth}},  {\em
  JHEP} {\bf 07} (2019) 095, [\href{http://arxiv.org/abs/1903.00462}{{\tt
  arXiv:1903.00462}}].

\bibitem{Takahashi:2019pqf}
F.~Takahashi and W.~Yin, {\it {QCD axion on hilltop by a phase shift of
  $\pi$}},  {\em JHEP} {\bf 10} (2019) 120,
  [\href{http://arxiv.org/abs/1908.06071}{{\tt arXiv:1908.06071}}].

\bibitem{Takahashi:2021tff}
F.~Takahashi and W.~Yin, {\it {Challenges for heavy QCD axion inflation}},
  {\em JCAP} {\bf 10} (2021) 057, [\href{http://arxiv.org/abs/2105.10493}{{\tt
  arXiv:2105.10493}}].

\bibitem{Takahashi:2013cxa}
F.~Takahashi, {\it {New inflation in supergravity after Planck and LHC}},  {\em
  Phys. Lett. B} {\bf 727} (2013) 21--26,
  [\href{http://arxiv.org/abs/1308.4212}{{\tt arXiv:1308.4212}}].

\bibitem{Handley:2020hdp}
W.~Handley and P.~Lemos, {\it {Quantifying the global parameter tensions
  between ACT, SPT and Planck}},  {\em Phys. Rev. D} {\bf 103} (2021), no.~6
  063529, [\href{http://arxiv.org/abs/2007.08496}{{\tt arXiv:2007.08496}}].

\bibitem{Mallaby-Kay:2021tuk}
M.~Mallaby-Kay et~al., {\it {The Atacama Cosmology Telescope: Summary of DR4
  and DR5 Data Products and Data Access}},  {\em Astrophys. J. Supp.} {\bf 255}
  (2021), no.~1 11, [\href{http://arxiv.org/abs/2103.03154}{{\tt
  arXiv:2103.03154}}].

\bibitem{Smith:2022hwi}
T.~L. Smith, M.~Lucca, V.~Poulin, G.~F. Abellan, L.~Balkenhol, K.~Benabed,
  S.~Galli, and R.~Murgia, {\it {Hints of Early Dark Energy in Planck, SPT, and
  ACT data: new physics or systematics?}},
  \href{http://arxiv.org/abs/2202.09379}{{\tt arXiv:2202.09379}}.

\end{thebibliography}\endgroup
\end{document}